\documentclass[showpacs]{revtex4}

\usepackage{amsfonts}
\usepackage[gnuplot36beta]{egplot}
\newcommand{\acro}[1]{#1}
\newcommand{\vect}[1]{\vec{#1}}

\begin{document}

\begin{egpfile}

\egpfigprelude{
  set term post enhanced eps dashed color size 7.5cm, 7.5cm lw 2 15;
  set pointsize 1.5;
  set size square;
}

\title{Giant non-equilibrium fluctuations in diffusion in freely-suspended liquid films.}
\author{Doriano Brogioli}
\affiliation{Energiespeicher- und Energiewandlersysteme, Universit\"at Bremen,
Wiener Stra{\ss}e 12, 28359 Bremen, Germany}
\author{Alberto Vailati}
\affiliation{Dipartimento di Fisica, Universit\`a degli Studi di Milano, I-20133 Milano, Italy}


\begin{abstract}
Experimental work has shown that non-equilibrium concentration
fluctuations arise during free diffusion in fluids and theoretical
analysis has been carried on.  The results show that, in usual
three-dimensional fluids, the phenomenon is extremely weak, in terms
of amplitude of the fluctuations and of corrugation of the diffusion
wave fronts. In this paper, we show that the phenomena strongly
depends on the dimensionality of the system: by extending the theory
to two dimensional systems, we show that the root mean square
amplitude of the fluctuations and the wave front corrugation become
much stronger. We also present an evaluation of the Hausdorf dimension
of the expected fluctuations.  Experimentally, two-dimensional liquid
systems can be realised as freely suspended liquid films; experiments
and theoretical works on diffusion in such systems showed that the
dynamics is deeply affected by the viscous drag exerted by the fluids
(e.g. air) surrounding the film. We provide an evaluation of the drag
on the fluctuations. In particular, we study the case of a
concentration profile that is initially gaussian: it can be directly
compared with the results from a fluorescence recovery after
photobleaching (\acro{FRAP}) experiment. We propose that this theory
and the related experiments can be relevant for describing the
diffusion along the cellular membranes of living organisms.
\end{abstract}

\maketitle

\section{Introduction}

Fick's flow arises whenever a concentration gradient is present in a
multicomponent fluid. The presence of this flux is the evidence of
non-equlibrium condition; fluctuations much stronger than equlibrium
ones have been reported in systems during diffusion between couples of
ordinary miscible fluids.~\cite{vailati97_2, vailati98, brogioli00,
  brogioli00_2} These non-equilibrium fluctuations are originated by
the termally excited velocity fluctuations, that displace volumes of
fluid in layers with different concentrations; it has been shown that
this effect accounts for the whole Fick's flow.~\cite{brogioli01}

Until now, the experiments were performed with usual three dimensional
fluids, where the fluctuations are very small~\cite{in_scrittura}: in
microgravity~\cite{gradflex} (a conditions under which they become
stronger) the fluctuations are less than 1/1000 of the macroscopic
concentration difference in the solution.  The fluctuations are
unlikely to produce any effect on chemical reactions or on physical
processes.  In order to investigate the phenomenon in different
dimensionalities, in Sect.~\ref{sect:mathematical:model}, we recall
the theory of non-equilibrium fluctuations~\cite{vailati97} and we
extend it to two-dimensional fluids.  The result is that the
fluctuations are much more intense and the long wavelenght
fluctuations significantly contribute to the diffusion process.

As already highlighted, the diffusive flow can be seen as the effect
of mixing due to the non-equilibrium
fluctuations~\cite{brogioli01}. In Sect.~\ref{sect:diffusion}, we show
that the contribution of the long wavelength fluctuations is
negligible in three-dimensional fluids and thus the non-equilibrium
fluctuations can be considered as the low-frequency tail of the
diffusion process, which mainly happens at the molecular level.  On
the other hand, the contribution of the long wavelength fluctuations
to the diffusion flow diverges for two-dimensional fluids, thus
emphasising that the long wavelength fluctuations play a major role in
two dimensions. The divergence has been known since a long time as the
so-called ``Stokes paradox''; here, the paradox is described in terms
of the fluctuating hydrodynamics theory.

The extension of the fluctating hydrodynamics to two dimensions also
allows us to evaluate the amplitude of the fluctuations (see
Sect.~\ref{sect:root:mean:square}). We find that the fluctuations
should be much stronger, of the order of 10\% of the macroscopic
concentration difference.

Two-dimensional fluid models have been used for describing
freely-suspended liquid films. The idea is that a thin, freely
suspended liquid layer should follow two-dimensional hydrodynamics on
length scales longer than the layer thickness.  Many experiments have
been performed on soap films~\cite{cheung96, rivera00, goldburg95_1,
  couder84, couder89}, that consist of two monolayers of amphiphilic
molecules around a layer of water. The layer thickness ranges from the
$4\mathrm{nm}$ of the Newton black film to many micrometers; features
down to a fraction of a millimeter should be considered two
dimensional \cite{goldburg95_1}.  Also liquid
crystals~\cite{bechhoefer97, other_liquid_crystal_films} can form
films, by a different effect. Smectic A spontaneously form a layered
structure, with the elongated molecules aligned to the surface
normal. When a film is drawn from this kind of liquid crystal, it is
formed by a set of monolayers, ranging from two to many hundreds
\cite{bechhoefer97}; the thickness can be selected with extreme
accuracy. Because of the absence of translational order in the
monolayers, the film behaves as a two dimensional system.  The
membranes of cells of living organisms~\cite{cells} can also be
considered as two-dimensional fluids, although their dynamics is
strongly affected by rigid structures inside them and by the
interaction with the surrounding fluids.

Various hydrodynamic phenomena have been observed in two-dimensional
systems, including turbulent and laminar flows~\cite{rivera00,
  goldburg95_1, goldburg96, couder84, couder89} and
convection~\cite{goldburg95_2}.  The experiments are discussed in
terms of two dimensional hydrodynamics. An important result is that
the role of the air surrounding the film cannot be neglected; in
turbulent flows, it damps the turbulent vortices. Rivera et
al. \cite{rivera00} phenomenologically modeled this damping by adding
a linear drag term in the two dimensional Navier-Stokes
equations. They found the dissipation due to air friction to be a
significant energy dissipation mechanism in their experimental system.

Other experiments addressed diffusion~\cite{bechhoefer97} and Brownian
motion~\cite{cheung96} in two-dimensional systems.  Also in these
cases, the effect of air friction can be seen. Neglecting air
friction, the mobility of a particle in a two dimensional system
should be infinite, that is, a particle subjected to a steady force
should accelerate indefinitely: this constitutes theso-called ``Stokes
paradox'', which would lead to an infinite diffusion
coefficient. However, the viscous drag exerted by the surrounding
fluids quenches the low-frequency velocity fluctuations and decreases
the mobility to a finite value, given by the well-known Saffman
formula.~\cite{saffman76}. The formula has been tested both in soap
films \cite{cheung96} and in liquid crystal films \cite{bechhoefer97}.

In Sects.~\ref{sect:diffusion} and \ref{sect:root:mean:square} we also
discuss the phenomenon of the non-equilibrium fluctuations in
freely-suspended liquid films taking into account the viscous drag of
the fluids surrounding the film. We show that the amplitude of the
fluctuations is extremely strong also in this case and that the long
wavelength fluctuations strongly contribute to the diffusion process.

In Sect.~\ref{sect:graphical:representation} we graphically compare
the fluctuations in three- and two-dimensional fluids and in liquid
films with viscous drag exerted by the surrounding gasses. We also
show synthetic images simulating the fluctuations in the three cases,
along with the iso-concentration curves representing the diffusion
wave fronts, which are corrugated due to the presence of the fluctuations.

It has been argued~\cite{gradflex} that the iso-concentration curves
are fractals; this conjecture has been proved to be
false~\cite{in_scrittura}.  In Sect.~\ref{sect:hausdorff:dimension} we
evaluate the fractal dimension of the iso-concentration curves in
two-dimensional fluids. We conclude that the iso-concentration curves
are not fractal neither in this case.

The results should be valid for completely liquid films. In the
conclusions, Sect.~\ref{sect:conclusions}, we propose that the
measurement of deviations from the expected spectrum could be used to
study the effect of rigid structures in the film, or the interaction
with surrounding media. Such information could be relevant for
obtaining information on the cellular membranes. Moreover, the motion
of receptors on the cellular membranes could be affected by the
non-equilibrium fluctuations.

\section{Mathematical model of the non-equilibrium fluctuations.}
\label{sect:mathematical:model}

We use the so-called ``fluctuating hydrodynamic'' approach~\cite{landau}.

The velocity correlation function for a two- and three-dimensional
liquid is:
\begin{equation}
\label{eq:velocity:correlation}
\left<
\vect{n}\cdot\vect{u}\left(\vect{q},\omega\right) \quad
\vect{n}\cdot\vect{u}^*\left(\vect{q'},\omega'\right)
\right>
=
\delta\left(\vect{q}-\vect{q}'\right)
\delta\left(\omega-\omega'\right)
\frac{K_BT\nu}{8\pi^4\rho}
\frac{q^2-\left(\vect{q}\cdot\vect{n}\right)^2}{\omega^2+\nu^2q^4}
\end{equation}
where $\vect{u}$ is the hydrodynamic velocity, $n$ is a generic unit
vector, and the wave-vector $q$ is in $\mathbb{R}^2$ or
$\mathbb{R}^3$ depending on the dimensionality of the system.
It sould be noticed that, in 2d, $\rho$ is the superficial mass
density.

The velocity correlation function for a freely suspended liquid film,
taking into account the viscous drag exerted by the surrounding
fluids, is the sum of two functions of $\omega$, with characteristic
decay times associated to the viscous damping in the liquid film and
in the surrounding fluids.~\cite{brogioli2005_eprint}. Here we report
only the value of the velocity correlation function for $\omega=0$:
\begin{equation}
\left<
\vect{n}\cdot\vect{u}\left(\vect{q},\omega=0\right) \quad
\vect{n}\cdot\vect{u}^*\left(\vect{q'},\omega'\right)
\right>
=
\delta\left(\vect{q}-\vect{q}'\right)
\delta\left(\omega'\right)
\frac{K_BT}{8\pi^4\rho \nu_0}
\frac{q^2-\left(\vect{q}\cdot\vect{n}\right)^2}{q^3\left(q_C+q\right)} ,
\label{eq:static:spectrum:2d}
\end{equation}
where the characteristic wave vector $q_C$ is:
\begin{equation}
q_C=\frac{2\eta_1}{h\eta_0},
\end{equation}
where $\eta_0$ and $\eta_1$ are the shear viscosities of the film and
the surrounding fluid, $\rho_0$ and $\rho_1$ are the volumetric
densities, $\rho$ is the surface density $\rho=\rho_0h$, and $h$ is the
thickness of the film.

The characteristic wave vector $q_C$ represents the wave vector below
which the velocity fluctuations are damped by the viscous drag of the
surrounding fluids.  For a soap film in air, the values of densities
are $\rho_0=10^3 \mathrm{Kg}/\mathrm{m}^3$ and
$\rho_1=1.3\mathrm{Kg}/\mathrm{m}^3$. The effective viscosity of the
film is $\nu_0=1.6\cdot10^{-6} \mathrm{m}^2/\mathrm{s}$,
\cite{rivera00}, evaluated by means of Trapeznikov relation
\cite{trapeznikov57}. The viscosity of air is
$\nu_1=1.43\cdot10^{-5}\mathrm{m}^2/\mathrm{s}$
\cite{handbook_chem_phys}.  From these values, for a $2\mu\mathrm{m}$
thick film, $q_C\approx 1.2\cdot10^4\mathrm{m}^{-1}$, corresponding to
a $0.5\mathrm{mm}$ wavelength.


The time evolution of the concentration $c\left(\vect{x},t\right)$ is given by 
convection and diffusion:
\begin{equation}
\frac{\partial}{\partial t}
c\left(\vect{x},t\right) 
= 
-\vect{u}\left(\vect{x},t\right)\cdot\vect{\nabla}c\left(\vect{x},t\right)
+D\nabla^2c\left(\vect{x},t\right)
\label{eq_concentrazione_spazio_reale}
\end{equation}
where $\vect{u}\left(\vect{x},t\right)$ is the velocity field and $D$
is the diffusion constant. The vectors can be in 3 or 2 dimensions,
depending on the physical system we want to describe.

We assume that the concentration gradient is a constant, since it is
always much more than the fluctuations.  In Fourier space:
\begin{equation}
-i\omega c\left(\vect{q},\omega\right) 
= 
-\vect{u}\left(\vect{q},\omega\right)\cdot\vect{\nabla}c_0
-q^2Dc\left(\vect{q},\omega\right)
\end{equation}
By solving with respect to the concentration we get:
\begin{equation}
c\left(\vect{q},\omega\right)
= 
-\frac{
\vect{u}\left(\vect{q},\omega\right)\cdot\vect{\nabla}c_0
}{
-i\omega + q^2D
}
\label{eq:solution:c}
\end{equation}
From this expression, we obtain the correlation functions:
\begin{equation}
\left<
c\left(\vect{q},\omega\right)
c^*\left(\vect{q'},\omega'\right)
\right>
=
\nabla c_0^2
\frac{\left<
\vect{\hat{z}}\cdot\vect{u}\left(\vect{q},\omega\right) \quad
\vect{\hat{z}}\cdot\vect{u}^*\left(\vect{q'},\omega'\right)
\right>}{
D^2q^4+\omega^2
}
\label{eq:c:spectrum:function:u}
\end{equation}
where $\vect{\hat{z}}$ is the unit vector pointing in the direction of
the concentration gradient, as the $z$ axis.

The static power spectrum of the fluctuations can be derived by
integrating over $\omega$. The calculation is performed assuming that
the diffusion time is much longer than the viscous time, since $D \ll
\nu$:
\begin{equation}
\left<
c\left(\vect{q},t\right)
c^*\left(\vect{q'},t\right)
\right>
=
\nabla c_0^2
\frac{\pi}{Dq^2}
\int
\left<
\vect{\hat{z}}\cdot\vect{u}\left(\vect{q},\omega=0\right) \quad
\vect{\hat{z}}\cdot\vect{u}^*\left(\vect{q'},\omega'\right)
\right>
\mathrm{d}\omega' 
\label{eq:c:static:spectrum:function:u}
\end{equation}

In the case of two- and three-dimensional fluids, we use
Eq.~\ref{eq:velocity:correlation} and we
obtain~\cite{vailati98,brogioli01}:
\begin{eqnarray}
\left<c\left(\vect{q},t\right)c^*\left(\vect{q}',t\right)\right>=
\delta\left(\vect{q}-\vect{q}'\right)
\frac{K_BT}{8\pi^3\rho}\nabla c^2
\frac{1}{\nu D q^4}
\frac{q^2-\left(\vect{q}\cdot\vect{\hat{z}}\right)^2}{q^2}
\label{eq:c:power:spectrum}
\end{eqnarray}

For the case of the freely suspended liquid film, with viscous drag of
the surrounding fluids, we use Eq.~\ref{eq:static:spectrum:2d} for
expressing the velocity correlation function needed by
Eq.~\ref{eq:c:static:spectrum:function:u}:
\begin{equation}
\left<c\left(\vect{q},t\right)c^2\left(\vect{q}',t\right)\right>=
\delta\left(\vect{q}-\vect{q}'\right)
\frac{K_BT}{8\pi^3\rho}
\nabla c_0^2
\frac{1}{\nu_0 D q^3 \left(q_C+q\right)}
\frac{q^2-\left(\vect{q}\cdot\vect{\hat{z}}\right)^2}{q^2}
\label{eq:c:power:spectrum:drag}
\end{equation}

\section{Fluctuations as the origin of Fick's flow}
\label{sect:diffusion}

In fluctuating hydrodynamic theory, the Fick's flow $\vect{\Phi}$ is
interpreted as a fluctuation of flow with a non-vanishing
average~\cite{brogioli01}:
\begin{equation}
\vect{\Phi} = \left< 
c\left(\vect{x}=0,t=0\right) 
\vect{u}\left(\vect{x}=0,t=0\right)
\right> .
\end{equation}
The only non-vanishing component of $\vect{\Phi}$ is along the
macroscopic concentration gradient $\nabla c$, directed as the unit
vector $\vect{\hat{z}}$. We express the fields in terms of their
Fourier transform:
\begin{equation}
\vect{\hat{z}}\cdot\vect{\Phi} = \int
\left< 
c\left(\vect{q},\omega\right) \quad
\vect{\hat{z}}\cdot\vect{u}^*\left(\vect{q'},\omega'\right)
\right> 
\mathrm{d}\vect{q}\mathrm{d}\vect{q'}
\mathrm{d}\omega \mathrm{d}\omega'
.
\end{equation}
We express $c\left(\vect{q},\omega\right)$ by using Eq.~\ref{eq:solution:c}:
\begin{equation}
\vect{\hat{z}}\cdot\vect{\Phi} = - \nabla c
\int
\frac{1}{-i\omega + q^2D}
\left< 
\vect{\hat{z}}\cdot\vect{u}\left(\vect{q},\omega\right)
\quad
\vect{\hat{z}}\cdot\vect{u}^*\left(\vect{q'},\omega'\right)
\right> 
\mathrm{d}\vect{q}\mathrm{d}\vect{q'}
\mathrm{d}\omega \mathrm{d}\omega'
.
\end{equation}
We rewrite the factor in the integral as:
\begin{equation}
\vect{\hat{z}}\cdot\vect{\Phi} = - \nabla c
\int
\frac{i\omega + q^2D}{\omega^2 + q^4D^2}
\left< 
\vect{\hat{z}}\cdot\vect{u}\left(\vect{q},\omega\right)
\quad
\vect{\hat{z}}\cdot\vect{u}^*\left(\vect{q'},\omega'\right)
\right> 
\mathrm{d}\vect{q}\mathrm{d}\vect{q'}
\mathrm{d}\omega \mathrm{d}\omega'
.
\end{equation}
The integral is a real number; this is consistent with the fact that the
imaginary part of the integrand is an odd function of $\omega$ that can be neglected:
\begin{equation}
\vect{\hat{z}}\cdot\vect{\Phi} = - \nabla c
\int
\frac{q^2D}{\omega^2 + q^4D^2}
\left< 
\vect{\hat{z}}\cdot\vect{u}\left(\vect{q},\omega\right)
\quad
\vect{\hat{z}}\cdot\vect{u}^*\left(\vect{q'},\omega'\right)
\right> 
\mathrm{d}\vect{q}\mathrm{d}\vect{q'}
\mathrm{d}\omega \mathrm{d}\omega'
.
\end{equation}
Now we assume that the time correlation of the velocity fluctuations
is much shorter than the diffusive time:
\begin{equation}
\vect{\hat{z}}\cdot\vect{\Phi} = - \nabla c \pi
\int
\left< 
\vect{\hat{z}}\cdot\vect{u}\left(\vect{q},\omega=0\right)
\quad
\vect{\hat{z}}\cdot\vect{u}^*\left(\vect{q'},\omega'\right)
\right> 
\mathrm{d}\vect{q}\mathrm{d}\vect{q'}
\mathrm{d}\omega'
.
\end{equation}
By comparing this expression with Fick's law:
\begin{equation}
D = \pi
\int^{\Lambda}_{Q} 
\left< 
\vect{\hat{z}}\cdot\vect{u}\left(\vect{q},\omega=0\right)
\quad
\vect{\hat{z}}\cdot\vect{u}^*\left(\vect{q'},\omega'\right)
\right> 
\mathrm{d}\vect{q}\mathrm{d}\vect{q'}
\mathrm{d}\omega'
.
\label{eq:generic:D:u}
\end{equation}
where we introduced the cut off wavevectors $\Lambda$, of the order of
$\pi/a$, where $a$ is the radius of the diffusing particle, and
$Q\approx \pi/L$, where $L$ is the macroscopic size of the fluid
system.

By using Eq.~\ref{eq:velocity:correlation} for calculating $D$ with
Eq.~\ref{eq:generic:D:u}:
\begin{equation}
D = \frac{K_BT}{8\pi^3\rho\nu} \int^{\Lambda}_{Q}
\frac{q^2-\left(\vect{q}\cdot\hat{z}\right)^2}{q^4}
\mathrm{d}\vect{q}
\end{equation}
For the three-dimensional case, we get:
\begin{equation}
D_{3D} = \frac{K_BT}{3\pi^2\rho\nu} 
\left(\Lambda - Q\right) 
\label{eq:diffusion:coefficient:3d:cutoff}
\end{equation}
and for the two-dimensional case:
\begin{equation}
D_{2D} = \frac{K_BT}{8\pi^2\rho\nu_0} 
\log\frac{\Lambda}{Q}
\label{eq:diffusion:coefficient:2d:cutoff}
\end{equation}

By using Eq.~\ref{eq:static:spectrum:2d} for calculating $D$ with
Eq.~\ref{eq:generic:D:u} for the case of the freely suspended liquid
film with viscous drag:
\begin{equation}
D = \frac{K_BT}{8\pi^3\rho \nu_0}
\int_Q^{\Lambda}{
\frac{q^2-\left(\vect{q}\cdot\hat{z}\right)^2}{q^3\left(q+q_C\right)}
\mathrm{d}\vect{q}}
.
\end{equation}
By integrating:
\begin{equation}
D_{VD} =
\frac{K_BT}{8\pi^2\rho \nu_0}
\ln\frac{\Lambda + q_C}{Q + q_C}
\label{eq:diffusion:coefficient:drag:cutoff}
\end{equation}

\begin{figure}
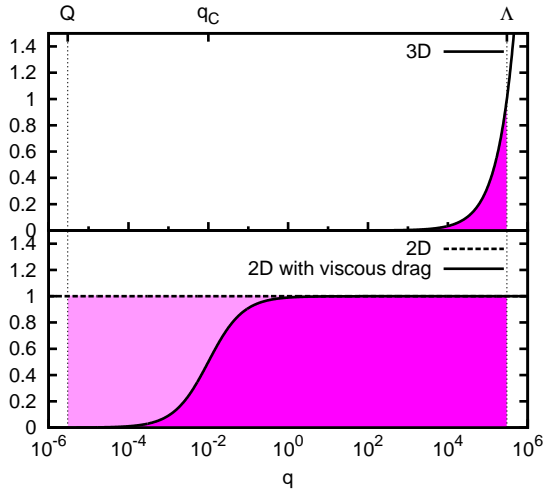

\begin{egp}
set size 1., 1.5
set size nosquare
set logscale xx2
Q=3e-6
Lambda=3e5
qC=0.01
f3D(q)=q/Lambda
f2D(q)=1.
f2DVD(q)=1./(1.+qC/q)
set arrow from Q,0. to Q,1.5 nohead lt 0 front
set arrow from Lambda,0. to Lambda,1.5 nohead lt 0 front
set samples 1000
set multiplot
set lmargin at screen 0.1
set rmargin at screen 0.95
#
set tmargin at screen 0.55
set bmargin at screen 0.2
set xtics nomirror
set format x "10^{
set xlabel "q"
plot [1e-6:1e6][0:1.5] \
  (x<Lambda && x>Q)?f2D(x):0/0 notitle with filledcurves x1 lc rgb "#ff99ff",\
  (x<Lambda && x>Q)?f2DVD(x):0/0 notitle with filledcurves x1 lc rgb "magenta",\
  f2D(x) title "2D" w l lt 2 lw 2 lc 0,\
  f2DVD(x) title "2D with viscous drag" w l lt 1 lw 2 lc 0
#
unset xlabel
set tmargin at screen 0.9
set bmargin at screen 0.55
set format x ""
set x2tics ("Q" Q, "{/Symbol L}" Lambda, "q_C" qC)
plot [1e-6:1e6][0:1.5]\
  (x<Lambda && x>Q)?f3D(x):0/0 notitle with filledcurves x1 lc rgb "magenta",\
  f3D(x) title "3D" w l lt 1 lw 2 lc 0
#
unset multiplot
\end{egp}
\caption{Graphical representation of the diffusion coefficient as the
  integral of the velocity power spectrum for $\omega=0$. The
  diffusion coefficient is the area below the curve, between the low-
  and high-wave length cut-offs $Q$ and $\Lambda$. 
  \label{fig:diffusion:coefficient}}
\end{figure}

The integrals leading to Eq.~\ref{eq:diffusion:coefficient:3d:cutoff},
\ref{eq:diffusion:coefficient:2d:cutoff} and
\ref{eq:diffusion:coefficient:drag:cutoff} are graphycally represented
in Fig.~\ref{fig:diffusion:coefficient} after changing the integration
variable to $\ln q$ instead of $q$. We can see that, in all the cases,
the integrals diverge for large values of the high-wave length cut-off
$\Lambda$; we set $\Lambda=\pi/a$, where $a$ is the radius of the
molecules. 

In the three-dimensional case, only the long wave length components
contribute to the diffusion. For example, from
Eq.~\ref{eq:diffusion:coefficient:3d:cutoff} we can see that the 90\%
of the integral from 0 to $\Lambda$ is reached by integrating over the
wave lengths from $\Lambda/10$ to $\Lambda$, i.e. by considering only
the lengthscales smaller than 10 times the molecular radius $a$: in
three-dimensions, diffusion is a process that mainly involves the
molecular length scales. We can thus take the limit for $Q\to 0$ of
Eq.~\ref{eq:diffusion:coefficient:3d:cutoff}; we get a value that is
of the same order of magnitude of the well-known Stokes-Einstein
formula~\cite{tanford}:
\begin{equation}
D_{3D}=\frac{K_BT}{6\pi\eta a}
\label{eq:diffusion:coefficient:3d}
\end{equation}

In the two-dimensional case, the integral diverges for $Q\to 0$. This
is the consequence of the so-called ``Stokes
paradox''~\cite{saffman76}, i.e. the divergence of the mobility of a
particle in a two-dimensional liquid film when the size of the film
becomes infinite. In this two-dimensional case, the long wave lengths
are not only relevant but even give a divergence. In order to handle
this case, we set the cut-off $Q=2\pi/L$, where $L$ is the macroscopic
size of the liquid film, in
Eq.~\ref{eq:diffusion:coefficient:2d:cutoff}:
\begin{equation}
D_{2D} = \frac{K_BT}{4\pi\rho\nu_0} 
\log\frac{L}{a}
\label{eq:diffusion:coefficient:2d}
\end{equation}

The viscous drag exerted by the fluids surrounding the liquid film
damps the velocity fluctuations at wave vectors shorter than $q_C$. In
this case, the macroscopic length scales up to $2\pi/q_C$ contribute to
the diffusion, that is no more a molecular process. Due to the
damping, it is however possible to take the limit $Q\to 0$ of
Eq.~\ref{eq:diffusion:coefficient:drag:cutoff}; we get a value that is
of the same order of magnitude of the well-known Saffman formula for
the diffusion coefficient of a particle on a liquid film
\cite{saffman76,cheung96,bechhoefer97}:
\begin{equation}
D_{VD} =
\frac{K_BT}{4\pi\rho \nu_0}
\ln\frac{2}{\gamma a q_C} ,
\label{eq:diffusion:coefficient:drag}
\end{equation}
where $\gamma \approx 0.577$ is the Euler-Mascheroni constant.

\section{Root mean square amplitude of the fluctuations.}
\label{sect:root:mean:square}

The root mean square value of the fluctuations of concentration is
calculated by integrating the power spectrum. It has been already
evaluated for the three-dimensional case~\cite{in_scrittura}. The
calculation requires the introduction of a low-wavelength cut-off
$Q$. We assume that $Q\approx 2\pi/L$, where $L$ is the length of the
region over which the concentration gradient is present; the
concentration gradient is thus $\nabla c=\Delta c /L$.  For the
three-dimensional fluids, the root-mean-square amplitude of the
fluctuations is~\cite{in_scrittura}:
\begin{equation}
\label{eq_c_rms_3d:D}
c_{rms}=
\Delta c
\sqrt{ \frac{K_BT}{3\pi^3\rho L \nu D} }
\end{equation}
This expression can be further elaborated by using the Stokes-Einstein
relation~\cite{tanford}, already expressed by
Eq.~\ref{eq:diffusion:coefficient:3d}, for writing $D$ in terms of the
viscosity $\nu$ and the radius $a$ of the diffusing
particles. I obtain:
\begin{equation}
\label{eq_c_rms_3d}
c_{rms} =
\Delta c \sqrt{\frac{2}{\pi}} \sqrt{\frac{a}{L}}
\end{equation}
We thus conclude that, for three-dimensional systems, the amplitude of
the fluctuations decreases as the macroscopic length scale $L$ of the
system increases, with respect to the molecular size $a$.  For $L$ of
the order of the mm and $a$ of the order of the nm, the fluctuation
amplitude is of the order of 1/1000 of the macroscopic concentration
difference across the concentration gradient. The corrugation of the
diffusion wavefront can be roughly evaluated as $c_{rms}/\nabla c
\propto \sqrt{aL}$: it's proportional to the geometric average between
the macroscopic length scale $L$ and the molecular size $a$, showing
that the phenomenon is actually ``mesoscopic'' in amplitude, i.e. its
amplitude is intermediate between the macroscopic and microscopic
length scale.

Now we extend the calculation of the root mean square value of the
fluctuations of concentration, to the two-dimensional case:
\begin{eqnarray}
c_{rms}^2&=&
\frac{K_BT}{8\pi^3\rho}\nabla c^2
\frac{1}{\nu D}
\int_0^{2\pi} d\theta
\int_Q^{\infty} q dq
\frac{\sin^2\left(\theta\right)}{q^4}\\
&=&
\frac{K_BT}{16\pi^2\rho}\nabla c^2
\frac{1}{\nu D}
\frac{1}{Q^2}
\end{eqnarray}
By defining the cut-off $Q=\pi/L$ as above: 
\begin{equation}
c_{rms}=
\Delta c
\sqrt{ \frac{K_BT}{8\pi^4\rho \nu D} }
\label{eq:c:rms:D}
\end{equation}
The parameter $L$ does not explicitly appear in this expression for
$c_{rms}$; however, $D$ depends on $L$, as explained in the
Sect.~\ref{sect:diffusion}. We use
Eq.~\ref{eq:diffusion:coefficient:2d} for expressing the diffusion
coefficient $D$ in Eq.~\ref{eq:c:rms:D}:
\begin{equation}
c_{rms}=
\frac{\Delta c}
{\pi \sqrt{ \log \frac{L}{a}} } .
\label{eq:c:rms}
\end{equation}
We see that, also in this case, the fluctuation amplitude vanishes as
the ratio $L/a$ tends to infinity. However, the dependence is through
a logarithm and is thus much slower than in the three-dimensional case
(Eq.~\ref{eq_c_rms_3d}). For example, taking the case considered
above, with $L$ of the order of the mm and $a$ of the order of the nm,
the two-dimensional fluctuation amplitude is of the order of 10\% of
the macroscopic concentration difference.

By approximating the corrugation of the diffusion wave fronts with 
$c_{rms}/\nabla c$, we get:
\begin{equation}
h_{rms}=
\frac{L}
{\pi \sqrt{ \log \frac{L}{a}} } .
\end{equation}
We see that the corrugation is a relevant fraction of the thickness of
the diffusion layer, e.g. of the order of 10\% under the
above-mentioned conditions.

Now we calculate the root mean square value of the fluctuations of
concentration for the freely suspended liquid film with the viscous
drag exerted by the surrounding fluids:
\begin{eqnarray}
c_{rms}^2&=&
\frac{K_BT}{8\pi^3\rho}\nabla c^2
\frac{1}{\nu_0 D}
\int_0^{2\pi} d\theta
\int_Q^{\infty} q dq
\frac{\sin^2\left(\theta\right)}{q^3\left(q_C+q\right)}\\
&=&
\frac{K_BT}{8\pi^2\rho}\nabla c^2
\frac{1}{\nu_0 D}
\left( \frac{1}{q_CQ} - \frac{1}{q_C^2} \log \frac{Q+q_C}{Q} \right)
\label{eq_c_rms_drag:D}
\end{eqnarray}
Here the diffusion coefficient $D$ must be calculated by means of the
Saffman formula Eq.~\ref{eq:diffusion:coefficient:drag}:
\begin{equation}
\label{eq_c_rms_drag}
c_{rms} = \Delta c
\frac{1}{\sqrt{2\pi}}
\frac{1}{L\sqrt{\ln\frac{2}{\gamma a q_C}}}
\sqrt{ \frac{L}{\pi q_C} - \frac{1}{q_C^2} \log \left( 1 + \frac{L q_C}{\pi} \right) }
\end{equation}
In the limit $Q\gg q_C$:
\begin{equation}
c_{rms} = \Delta c
\frac{1}{2\sqrt{\pi^3}}
\frac{1}{\sqrt{\ln\frac{2}{\gamma a q_C}}}
\end{equation}
which has a value close to that of Eq.\ref{eq:c:rms}. In the limit 
$Q\ll q_C$:
\begin{equation}
c_{rms} = \Delta c
\frac{1}{\sqrt{2\pi}}
\frac{1}{\sqrt{\ln\frac{2}{\gamma a q_C}}}
\sqrt{ \frac{1}{L \pi q_C} }
\end{equation}
which has a $1/\sqrt{L}$ dependence similar to Eq.~\ref{eq_c_rms_3d}.

\section{Graphical representation of the fluctuations}
\label{sect:graphical:representation}

\begin{figure}
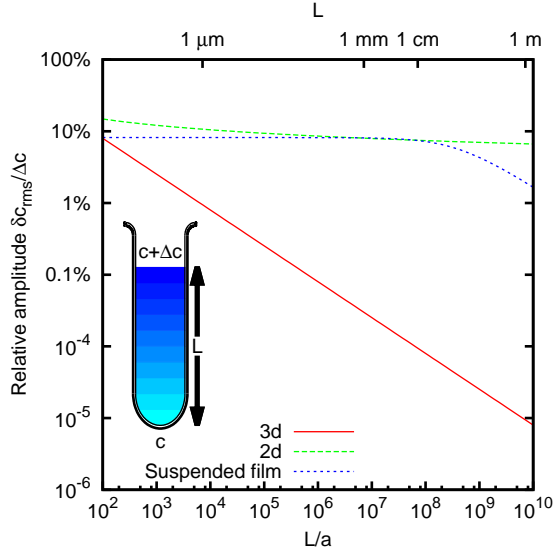

\begin{egp}
set multiplot
#
set xlabel "L/a"
set ylabel "Relative amplitude {/Symbol d}c_{rms}/{/Symbol D}c"
set x2label "L"
set logscale xx2y
a=0.14e-9
qc=100.
set xtics nomirror
set key bottom left
set format x "10^{
set ytics ("100
set x2tics ("1 {/Symbol m}m" 1e-6/a, "1 mm" 1e-3/a, "1 cm" 0.01/a, "1 m" 1./a)
plot [100:1e10][1e-6:1] \
  sqrt(2./pi)/sqrt(x) title "3d",\
  1./(pi*sqrt(log(x))) title "2d",\
  (a*x*qc<0.0001)? 4./sqrt(2.*pi)/(a*x)/sqrt(log(2/0.57/a/qc))*sqrt(a**2*x**2/pi**2/2.) : 4./sqrt(2.*pi)/(a*x)/sqrt(log(2/0.57/a/qc))*sqrt(a*x/qc/pi-log(1+a*x*qc/pi) /qc**2) title "Suspended film"
#
set size nosquare
unset logscale
unset xlabel
unset ylabel
set border 0
unset xtics
unset ytics
unset x2label
set lmargin at screen 0.1
set rmargin at screen 0.45
set tmargin at screen 0.63
set bmargin at screen 0.18
load "test_tube.gp"
#
unset multiplot
\end{egp}
\caption{Root-mean-square amplitude of the non-equilibrium
  fluctuations of concentration in three- and two-dimensional fluids,
  and in freely suspended liquid films surrounded by less viscous
  fluids. The amplitude is represented as a fraction of the
  macroscopic concentration difference $\Delta c$ across the layer of
  thickness $L$ over which diffusion takes place. The scale on the
  upper border assumes $a=0.14$~nm, the effective radius of water
  molecules. The value of $q_C$ is 100/m.
\label{fig:c:rms:comparison}}
\end{figure}

Figure~\ref{fig:c:rms:comparison} shows a comparison between the
root-mean-square amplitude of the non-equilibrium fluctuations of
concentration in three- and two-dimensional fluids, and in freely
suspended liquid films surrounded by less viscous fluids.  It can be
clearly seen that the fluctuations have a negligible amplitude in
three-dimensional fluids, when the macroscopic concentration gradient
extends over a thickness of the order of millimeters; in the same
conditions, the fluctuations are quite intense in the two-dimensional
fluids. The effect of the viscous drag exerted by the surrounding
fluids is to decrease the amplitude of the fluctuations when the size
of the film becomes larger than the wavelength corresponding to the
characteristic wave vector $q_C$.

\begin{figure}
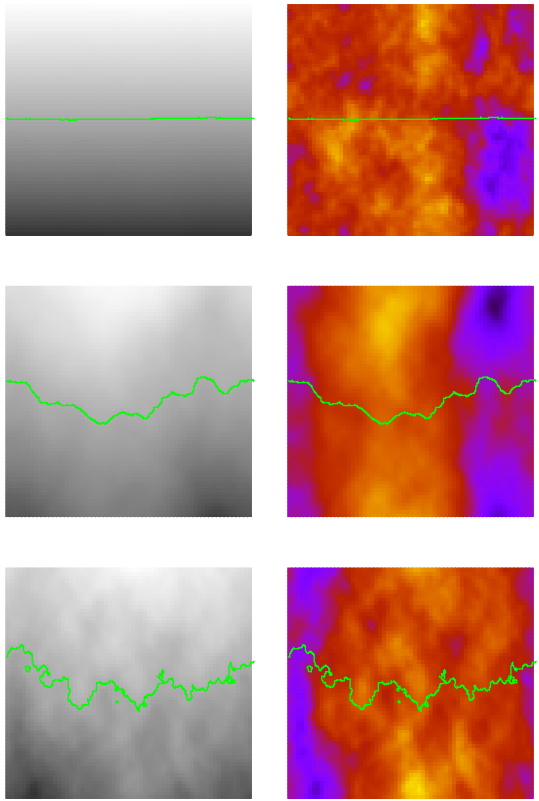

\begin{egp}
  set term post enhanced eps dashed color size 7.5cm, 11.25cm lw 2 15
  set multiplot layout 3,2
  set border 0
  unset xtics
  unset ytics
  unset ztics
  set view 0,0,2
  unset colorbox
  taglia(x)=(x>1500)?1500:((x<-500)?-500:x)
#
  set palette grey
  splot [][][-500:1500] "< gawk 'NF>0 {print $0; a=0}; NF==0 {a++; if (a==2) printf \"\\n\"; if (a>=2) printf \"\\n\"}' images.txt" matrix i 0 every 4:4 u 1:2:(taglia($3)) notitle w pm3d,\
     "images.txt" i 2 u 1:2:(0.) notitle w l lt 1 linecolor rgb "green"
#
  set palette color
  splot [][][-4000:4000] "< gawk 'NF>0 {print $0; a=0}; NF==0 {a++; if (a==2) printf \"\\n\"; if (a>=2) printf \"\\n\"}' images.txt" matrix i 1 every 4:4 notitle w pm3d,\
     "images.txt" i 2 u 1:2:(0.) notitle w l lt 1 linecolor rgb "green"
#
  set palette grey
  splot [][][-500:1500] "< gawk 'NF>0 {print $0; a=0}; NF==0 {a++; if (a==2) printf \"\\n\"; if (a>=2) printf \"\\n\"}' images.txt" matrix i 3 every 4:4 u 1:2:(taglia($3)) notitle w pm3d,\
     "images.txt" i 5 u 1:2:(0.) notitle w l lt 1 linecolor rgb "green"
#
  set palette color
  splot [][][-4000:4000] "< gawk 'NF>0 {print $0; a=0}; NF==0 {a++; if (a==2) printf \"\\n\"; if (a>=2) printf \"\\n\"}' images.txt" matrix i 4 every 4:4 notitle w pm3d,\
     "images.txt" i 5 u 1:2:(0.) notitle w l lt 1 linecolor rgb "green"
#
  set palette grey
  splot [][][-500:1500] "< gawk 'NF>0 {print $0; a=0}; NF==0 {a++; if (a==2) printf \"\\n\"; if (a>=2) printf \"\\n\"}' images.txt" matrix i 6 every 4:4 u 1:2:(taglia($3)) notitle w pm3d,\
     "images.txt" i 8 u 1:2:(0.) notitle w l lt 1 linecolor rgb "green"
#
  set palette color
  splot [][][-4000:4000] "< gawk 'NF>0 {print $0; a=0}; NF==0 {a++; if (a==2) printf \"\\n\"; if (a>=2) printf \"\\n\"}' images.txt" matrix i 7 every 4:4 notitle w pm3d,\
     "images.txt" i 8 u 1:2:(0.) notitle w l lt 1 linecolor rgb "green"
#$
  unset multiplot
\end{egp}
\caption{ Section of the fluid slab where diffusion takes place.  The
  ratio between the slab thickness and the molecular radius is $L/a =
  10^5$. For each row, the image on the left is the concentration
  profile, represented in grayscale; the right image represents the
  displacement of the concentration with respect to the
  constant-gradient macroscopic concentration
  profile. Iso-concentration curves, i.e. the diffusion wave-fronts,
  are also shown. Top row: three-dimensional fluid. Center row:
  two-dimensional fluid. Bottom row: freely suspended fluid film with
  viscous drag of the surrounding fluids.
\label{fig:simulation}}
\end{figure}

Figure~\ref{fig:simulation} shows simulations of the non-equilibrium
fluctuations. 

For the three-dimensional case, we start from
Eq.~\ref{eq:c:power:spectrum} and we evaluate the concentration
correlation along a section with $y=0$:
\begin{equation}
\left<c\left(q_x,y=0,q_z,t\right)c^*\left(q_x',y=0,q_z',t\right)\right>
=
\int
\left<c\left(q_x,q_y,q_z,t\right)c^*\left(q_x',q_y',q_z',t\right)\right>
\mathrm{d}q_y \mathrm{d}q_y'
\end{equation}
We get:
\begin{equation}
\left<c\left(q_x,y=0,q_z,t\right)c^*\left(q_x',y=0,q_z',t\right)\right>
=
\delta\left(q_x-q_x'\right) \delta\left(q_z-q_z'\right)
\frac{K_BT}{64\pi^2\rho}\nabla c^2
\frac{1}{\nu D}
\frac{4q_x^2+q_z^2}{\left(q_x^2+q_z^2\right)^{5/2}}
\end{equation}

For the two-dimensional case and for the freely suspended film with
viscous drag of the surrounding fluid, Eqs.~\ref{eq:c:power:spectrum}
and \ref{eq:c:power:spectrum:drag} directly give the correlation
function of the fluctuations that can be observed on the film.

In Fig.~\ref{fig:simulation} we can clearly notice that the
fluctuations in the three dimensional case are nearly invisible, while
they are clearly visible in the two other cases. We can also notice
that the iso-concentration curves are much more smooth in the
two-dimensional fluid.

\section{Hausdorff dimension of the diffusion wave fronts}
\label{sect:hausdorff:dimension}

It has been argued that the iso-concentration curves in
three-dimensional fluids are fractals because they are
self-similar~\cite{gradflex}. This has been recently proved to be
false~\cite{in_scrittura}. In this section, we analyse the case of the
two-dimensional fluid, in which the fluctuations appear to be much
stronger; we neglect the viscous drag of the surrounding fluids
because it would introduce a characteristic length scale $\/q_C$ thus
impeding the existence of a scaling laws.

We first consider two points inside the solution, displaced along or
perpendicularly with respect to the macroscopic concentration
gradient, and we evaluate the root mean square value of the
concentration difference between them. We call the two quantities
$\delta c_{\parallel}\left(\delta z\right)$ and $\delta
c_{\perp}\left(\delta x\right)$:
\begin{eqnarray}
\delta c_{\parallel}\left(\delta z\right)
&=& \sqrt{
\left<\left[
c\left(0,t\right) -
c\left(\vect{\hat{z}} \delta z,t\right)
\right]^2\right> }
\\
\delta c_{\perp}\left(\delta x\right)
&=& \sqrt{
\left<\left[
c\left(0,t\right) -
c\left(\vect{\hat{x}} \delta x,t\right)
\right]^2\right> }
\end{eqnarray}
where $\vect{\hat{x}}$ and $\vect{\hat{z}}$ are the unit vectors
perpendicular and parallel to the concentration gradient.

We will see that the integrals leading to such quantities do not
diverge: hence they are a better local characterisation of the
fluctuations than the root mean square value.

In polar coordinates:
\begin{eqnarray}
\delta c_{\parallel}^2\left(\delta z\right)
&=&
\frac{K_BT}{8\pi^3\rho}\nabla c^2
\frac{1}{\nu D}
\int_0^{2\pi} \mathrm{d}\theta
\int_0^{\infty} q \mathrm{d}q
\frac{\sin^2\left(\theta\right)}{q^4}
\left[2- 2\exp{-i q \delta z \cos \theta} \right]
\\
\delta c_{\perp}^2\left(\delta x\right)
&=&
\frac{K_BT}{8\pi^3\rho}\nabla c^2
\frac{1}{\nu D}
\int_0^{2\pi} \mathrm{d}\theta
\int_0^{\infty} q \mathrm{d}q
\frac{\sin^2\left(\theta\right)}{q^4}
\left[2- 2\exp{-i q \delta x \sin \theta} \right]
\end{eqnarray}

The integrals cannot be easily calculated explicitly. By changing the
integration variable to $t=q\delta x$ and $t=q\delta z$, we get:
\begin{eqnarray}
\label{eq:c:par:perp}
\delta c_{\parallel}\left(\delta z\right)
&=&
\sqrt{ \frac{K_BT}{8\pi\rho \nu D} }
\nabla c
\delta z \Gamma_{\parallel}
\\
\delta c_{\perp}\left(\delta x\right)
&=&
\sqrt{ \frac{K_BT}{8\pi\rho \nu D} }
\nabla c
\delta x \Gamma_{\perp}
\end{eqnarray}
where:
\begin{eqnarray}
\Gamma_{\parallel}
&=&
\frac{1}{\pi}
\sqrt{
\int_0^{2\pi}  \mathrm{d}\theta
\int_0^{\infty} \mathrm{d}t
\frac{\sin^2\left(\theta\right)}{t^3}
\left[2- 2\exp{-i t \cos \theta} \right]
} 
\\
\Gamma_{\perp}
&=&
\frac{1}{\pi}
\sqrt{
\int_0^{2\pi}  \mathrm{d}\theta
\int_0^{\infty} \mathrm{d}t
\frac{\sin^2\left(\theta\right)}{t^3}
\left[2- 2\exp{-i t \sin \theta} \right]
} 
\end{eqnarray}
The integrals converge and can be easily calculated numerically; we
obtain the values $\Gamma_{\parallel} \approx 0.56$ and
$\Gamma_{\perp} \approx 0.95$. From Eqs.~\ref{eq:c:par:perp}, we can conclude that the fluctuations
are slightly more extended in the direction perpendicular to the
macroscopic concentration gradient. 

Now we evaluate the corrugation of the iso-concentration curves. As above, we approximate the corrugation $h\left(\delta x\right)=c_{\perp}\left(\delta x\right)/\nabla c$. We get:
\begin{equation}
h\left(\delta x\right)
=
\sqrt{ \frac{K_BT}{8\pi\rho \nu D} }
 \Gamma_{\perp}
\delta x
\end{equation}
The dependence between $h$ and $\delta x$ is linear: $h\propto \delta
x$. In this case, a Hausdorff dimension 1 is expected~\cite{falconer,
  takayasu}. This linear dependence is connected with smoothness,
that, in this case, can be expressed by the Lipschitz condition; in
Fig.~\ref{fig:simulation} we see that the iso-concentration curves are
quite smooth. We can argue that the iso-concentration curves are not
fractal either in the two-dimensional case.

\section{Conclusions}
\label{sect:conclusions}

Our result allow to conclude that the non-equilibrium fluctuations
have a negligible amplitude in three-dimensional fluids, when the
macroscopic concentration gradient extends over a thickness of the
order of millimeters; in the same conditions, the fluctuations are
quite intense in the two-dimensional fluids.

The above results should be valid for completely liquid
films. However, important liquid films include the cellular membranes,
in which rigid structures are known to exist.  We propose that the
measurement of deviations from the expected power spectrum could be used to
study the effect of rigid structures in the film, or the interaction
with surrounding media. Such information could be relevant for
obtaining information on the cellular membranes. 

Moreover, we propose that the non-equilibrium fluctuations can affect
the motion of receptors on the cellular membranes.

\acknowledgments

\bibliography{2d.bib}

\newpage

%

%

%

%

%

\end{egpfile}
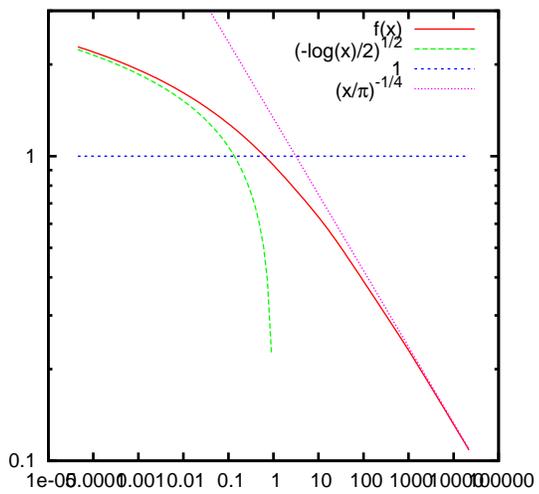
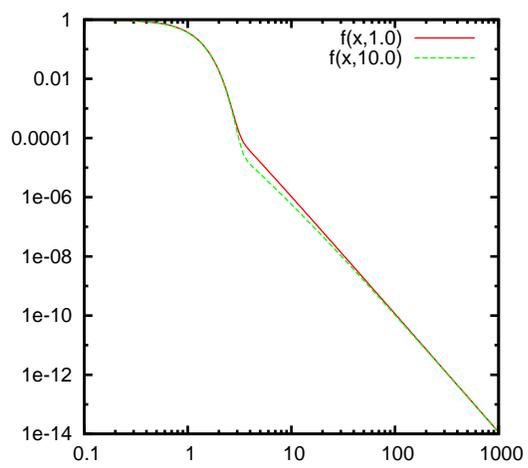
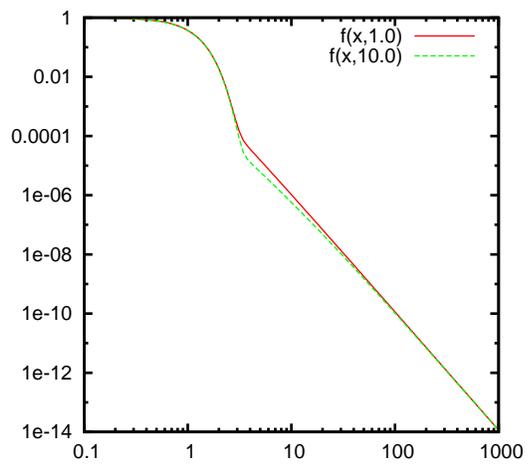
\end{document}